\documentclass[twocolumn,prl]{revtex4-1}
\usepackage{epsfig,amssymb,amsmath,epstopdf}
\begin{document}

\title{Shapiro Step at Nonequilibrium Conditions}
\author{Yu. M. Shukrinov~$^{1,2}$}
\author{M. Nashaat~$^{1,3}$}
\author{K. V. Kulikov~$^{1,2}$}
\author{ R. Dawood~$^{1,3}$}
\author{H. El Samman$^{4}$}
\author{Th. M. El Sherbini,~$^{3}$}
\address{$^{1}$ BLTP, JINR, Dubna, Moscow Region, 141980, Russia \\
$^{2}$ Dubna International University of Nature, Society, and Man, Dubna, Moscow region, 141980 Russia \\
$^{3}$Department of Physics, Cairo University, Cairo, Egypt\\
$^{4}$Department of Physics, Faculty of Science, Menoufiya University, Egypt}
\date{\today}

\begin{abstract}
Detailed numerical simulations of intrinsic Josephson junctions of high temperature superconductors under external electromagnetic radiation are performed taking into account a charge imbalance effect. We demonstrate that the charge imbalance is responsible for a slope in the Shapiro step the value of which increases with a nonequilibrium parameter. Coupling between junctions leads to the distribution of the slope's values along the stack. The nonperiodic boundary conditions shift the Shapiro step in IV-characteristics from the canonical position determined by  $V_{ss}=\hbar f /(2e)$, where $f$ is a frequency of external radiation. This fact makes ambiguous the interpretation of the experimentally found Shapiro step shift by the charge imbalance effect.
\keywords{Charge imbalance effect, microwave radiation, nonequilibrium,  Shapiro step, intrinsic Josephson junctions.}
\end{abstract}
\maketitle
The nonequilibrium effects created by stationary current injection in layered superconducting materials have been studied very intensively in recent years \cite{Artemenko97,Preis98,Shafranjuk99, Helm00,Helm01,Helm02,Bulaevskii02}. Actually, a  system of intrinsic Josephson junctions (IJJ) in high temperature superconductors cannot be in the equilibrium state at any value of the electrical current \cite{Koyama96,Dmitry}. The influence of charge coupling on Josephson plasma oscillations was stressed in Refs.\cite{Koyama96, Helm02}. However,  the charge imbalance in the systematic perturbation theory is considered only indirectly as far as it is induced by fluctuations of the scalar potential \cite{Artemenko97, Preis98,Helm01}. In Ref.\cite{keller05}, it is taken into account as an independent degree of freedom and, therefore, the results are different from those of earlier treatments.

Clear experimental evidence of the nonequilibrium effects in IJJ, which were explained by the charge imbalance in the superconducting layers produced by the quasiparticle current, was observed in Ref.\cite{rother03}. The experiments were based on the idea that the bias current generates charge accumulation on the layers between a resistive and superconducting junctions. The current through resistive junction is carried mostly by quasiparticles, while the current through a barrier in the superconducting state is carried by Cooper-pairs. It leads to charge fluctuations of the superconducting condensate in S-layers, which can be expressed by a shift of the chemical potential of the condensate and the charge imbalance between electron-like and hole-like quasiparticles. The authors of Ref.\cite{rother03} observed experimentally a shift of the Shapiro step (SS) voltage from the canonical value $V_{ss}=\hbar f /(2e)$ for the single mesa structures. They also detected an influence of the current through one mesa on the voltage measured on the other one in the double mesa structures. The results were also explained by the charge imbalance effect.

The answer to the question how strong the nonequilibrium effects are in the system is very important for different applications. Here we suggest a way to answer it. We study the nonequilibrium effects created by current injection in a stack of IJJ  under external electromagnetic radiation. The current-voltage characteristics (IV-characteristics) of IJJ are numerically calculated using the resistively and capacitively shunted junction model. The model takes into account the coupling between the layers and the quasiparticle charge imbalance effect \cite{Koyama96,sm-prl2007}.  We solve numerically a full set of  equations which include the first order differential equations for phase  differences, generalized Josephson relations and the kinetic equations. The boundary conditions based on the proximity effect are used. We obtain the branch structure of IV-characteristics and investigate the SS at different boundary conditions and nonequilibrium conditions.

\begin{figure}[!ht]
 \centering
\includegraphics[width=1\linewidth, angle =0]{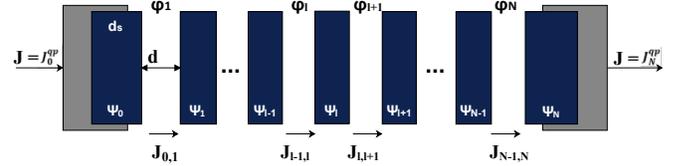}
\caption{ Layered system of $N+1$ superconducting layers forms a stack of Josephson junctions. Since the 0-th and N-th layers are in contact with normal metal, their thicknesses $d^0_s$ and $d^N_s$  are different from the thickness of the other S-layers $d_s$ inside of the stack due to the proximity effect.}
 \label{fig:layer}
\end{figure}

A system of $N+1$ superconducting layers (S-layers) presented in Fig. \ref{fig:layer} is characterized by the order parameter $\Delta_l(t) = \vert \Delta \vert \exp(i\theta_l(t))$ and time-dependent phase $\theta_l(t)$. The phase dynamics of the stack is described by a gauge invariant phase difference between the $S$-layers $ \varphi_l(t) \equiv \varphi_{l,l-1}(t)= \theta_{l}(t)-\theta_{l-1}(t)-\frac{2e}{\hbar}\int^{l}_{l-1}dz A_{z}(z,t) $, where $A_z(z,t)$ is the vector potential in the barrier \cite{keller05}. The thickness of the S-layer is comparable with the Debay screening length that leads to the generalized Josephson relation $\frac{d\varphi_l(t)}{dt} = \frac{2e}{\hbar} \Bigl( V_{l}(t) + \Phi_{l}(t) - \Phi_{l-1}(t)\Bigr)$ with a voltage $V_l$ between the layers $l-1$ and $l$, $ V_l(t) \equiv V_{l,l-1}(t)= \int_{l-1}^{l}dz E_z(z,t) $ and the gauge invariant scalar potential $\Phi_l(t)$ of the $S$-layer $\Phi_l(t)= \phi_l(t) - \frac{\hbar}{2e} \dot \theta_{l},  \label{Phi}$ where $\phi_l(t)$ is the electrical scalar potential. In contrast with the usual Josephson relation, the frequency of the Josephson oscillations is determined by $V_l$ and $\Phi_{l} - \Phi_{l-1}$. So in the nonequilibrium case the total energy $\hbar \dot \varphi_l(t)$ required to transfer a Cooper pair from the $l-1$ to $l$   is different from the equilibrium case by $\Phi_{l} - \Phi_{l-1}$ \cite{keller05}.

The nonperiodic boundary conditions (BCs) are characterized by the parameter $\gamma$ and, as we see below, the equations for the first and the last S-layers are different from the equation for the middle S-layer \cite{Koyama96,Matsumoto99}. The total current density $J_{l-1,l} \equiv J_{l}$ through each S-layer is given as a sum of displacement, superconducting, quasiparticle and diffusion terms:

\begin{eqnarray}
 	J_{l}=C\frac{dV_{l}}{dt}+J_{c}\sin\varphi_{l}+\frac{\hbar}{2eR}\dot{\varphi}_{l}+\frac{\Psi_{l-1}-\Psi_{l}}{R},
\end{eqnarray}

where $C$ is the capacitance, $J_{c}$ is the critical current density, and $R$ is the junction resistance. This equation together with the generalized Josephson relation and kinetic equations for $\Psi_l$

\begin{eqnarray}\label{current}
 	\frac{\partial\Psi_l}{\partial t}= \frac{4\pi r_D^2}{d_s^i} (J_{l}^{qp}-J_{l-1}^{qp})-\frac{\Psi_i}{\tau_{qp}}
\end{eqnarray}

describe the physics of IJJs in HTSC. In formula (\ref{current}),  $r_{D}$ is the Debye length, $d^{i}_{s}$ is the thickness of the S-layers, and $\tau_{qp}$ is the quasiparticle relaxation time.

In the dimensionless form the system of equations are

\begin{eqnarray}
 	\dot{v}_{l}&=&  \bigg{[} I -\sin\varphi_{l}-\beta\dot{\varphi}_{l} + A \sin{\omega} \tau + I_{noise} \nonumber \\ &+&\psi_{l}-\psi_{l-1}\bigg{]} \\
 	\dot{\varphi}_{1} &=& v_{1} - \alpha (v_{2}-(1+\gamma)v_{1}) +\dfrac{\psi_{1}-\psi_{0}}{\beta} \\
 	\dot{\varphi}_{l} &=& (1+2\alpha)v_{l} - \alpha (v_{l-1}+v_{l+1}) +\dfrac{\psi_{l}-\psi_{l-1}}{\beta} \\
 	\dot{\varphi}_{N}&=& v_{N} - \alpha (v_{N-1}-(1+\gamma)v_{N}) +\dfrac{\psi_{N}-\psi_{N-1}}{\beta} \\
 	\zeta_{0} \dot{\psi}_{0} &=& \eta_{0}  \left( I - \beta \dot{\varphi}_{0,1} + \psi_{1}-\psi_{0}\right) -\psi_{0} \\
 	\zeta_{l} \dot{\psi}_{l}&=& \eta_{l} ( \beta[\dot{\varphi}_{l-1,l} - \dot{\varphi}_{l,l+1} ]+ \psi_{1-1}+\psi_{l+1} - 2\psi_{l})  - \psi_{l}\nonumber  \\ \\
 	\zeta_{N} \dot{\psi}_{N} &=& \eta_{N}  \left( -I + \beta\dot{\varphi}_{N-1,N} + \psi_{N-1}-\psi_{N}\right) -\psi_{N} \hspace{0.3cm}
 	\label{eq:8}
 \end{eqnarray}

where the dot shows a derivative with respect to $\tau=\omega_pt$, $I=J/J_{c}$ is the dimensionless current, $\omega_p=\sqrt{\frac{2eJ_c}{\hbar C}}$ is plasma frequency and $\alpha = \epsilon\epsilon_{o} /2e^{2} N(0) d $ is the coupling parameter, $\epsilon$ is the dielectric constant, $\epsilon_{o}$ is the vacuum permittivity, $d$ is the distance between the superconducting layers and $N(0)$ is the density of states. Other dimensionless parameters are the dissipation parameter $\beta=\frac{\hbar\omega_p}{2eRI_c}$, the normalized quasiparticle relaxation time $\zeta_{l}=\omega_p \tau_{qp}$, the nonequilibrium parameter $\eta_{l}=\frac{4\pi r_D^2\tau_{qp}}{d^{l}_{s}R}$. The parameter of the nonperiodic boundary conditions $\gamma$ is $\gamma=\frac{d_s}{d_s^0}=\frac{d_s}{d_s^n}$. The term $A \sin \omega \tau$ introduces the effect of external radiation with amplitude A and frequency $\omega$, which are normalized to $J_{c}$ and $\omega_{p}$, respectively. To reflect the experimental situation, we have added the noise $I_{noise}$ in the bias current with the amplitude $\sim 10^{-8}$ which is produced by random number generator and its amplitude is normalized to the critical current density value $J_{c}$.

This system of equations is solved numerically using the fourth order Runge-Kutta method. We assume here that the nonequilibrium parameters for all the S-layers are the same (i.e., $\eta_{0} = \eta_{l}=\eta_{N}=\eta$). We consider the underdamped case with the McCumber parameter $\beta_{c}=25$ or $\beta=0.2$. In our simulations, we use the external radiation frequency $\omega =6$ to have SS on the outermost branch where all IJJs are in the rotating state, and we put the amplitude $A=1.6$ for a clear manifestation of the SSs. The method of simulations is described in detail in Ref.\cite{sg-prb2011}.

\begin{figure}[h!]
	\includegraphics[width=.9\linewidth, angle =-90]{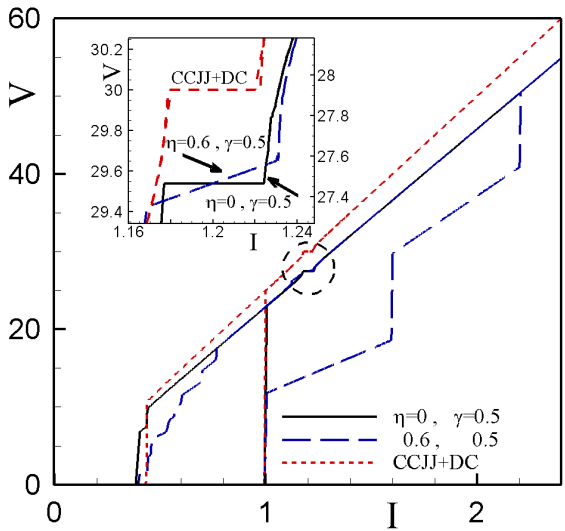}
	\caption {(Color online) The IV-characteristics of JJ stacks without the charge imbalance $\eta=0$ (solid line) and at $\eta=0.6$ (dashed line). The  results for the CCJJ+DC model (small dashed line) are shown for comparison.  The enlarged parts of the IV-characteristics with the SS are shown in the inset.}
	\label{fig:2}
\end{figure}

The coupled Josephson junctions at the nonequilibrium conditions are described by IV-characteristics with intensive branching near the critical current and in the hysteresis region, related to the transitions between the rotating and oscillating states of junctions in the stack \cite{Matsumoto99,SM-PhysC2006,sm-physc2006}. External radiation leads to the appearance of the Shapiro steps in IV-curve and a decrease in the hysteresis. In our study, we choose the values of radiation frequency $\omega=6$ and its amplitude A=$1.6$ to have a clear manifestation of the SS on the outermost branch in the middle of the hysteresis region. We consider here the underdamped junctions with the dissipation parameter $\beta=0.2$ and use the nonperiodic boundary conditions to reflect a proximity effect on the boundary between the normal electrode and superconducting layer. The nonperiodic boundary conditions are determined by the parameter $\gamma$ which demonstrates an effective change of the superconducting layer thickness near the electrode.  The simulated IV-characteristics of JJs stack in the case without the charge imbalance $\eta=0$ (solid line) and at $\eta=0.6$ (dashed line) are presented in Fig.\ref{fig:2}. The simulations have been made for the stacks with five JJs, coupling parameter $\alpha=0.5$ and $\gamma=0.5$. The IV-curve without the charge imbalance (in the CCJJ+DC model) at periodic boundary conditions are shown as well. We see that the position of the SS (dashed line) in the IV-characteristics corresponds to the canonical value of the SS voltage  $V=30$ in agreement with the value of external frequency  $\omega=6$  and a number of junctions in the stack $N=5$. The nonperiodic boundary conditions with  $\gamma\neq0$ shift the outermost branch relatively to the curve of the CCJJ+DC model, leading to the corresponding shift of the Shapiro steps. \emph{The charge imbalance manifests itself as appearance of the slope in the Shapiro step, which is clearly demonstrated in the inset for the case $\eta=0.6$.}
\begin{figure}[h!]
	\centering
		\includegraphics[ width=0.4\linewidth, angle =-90]{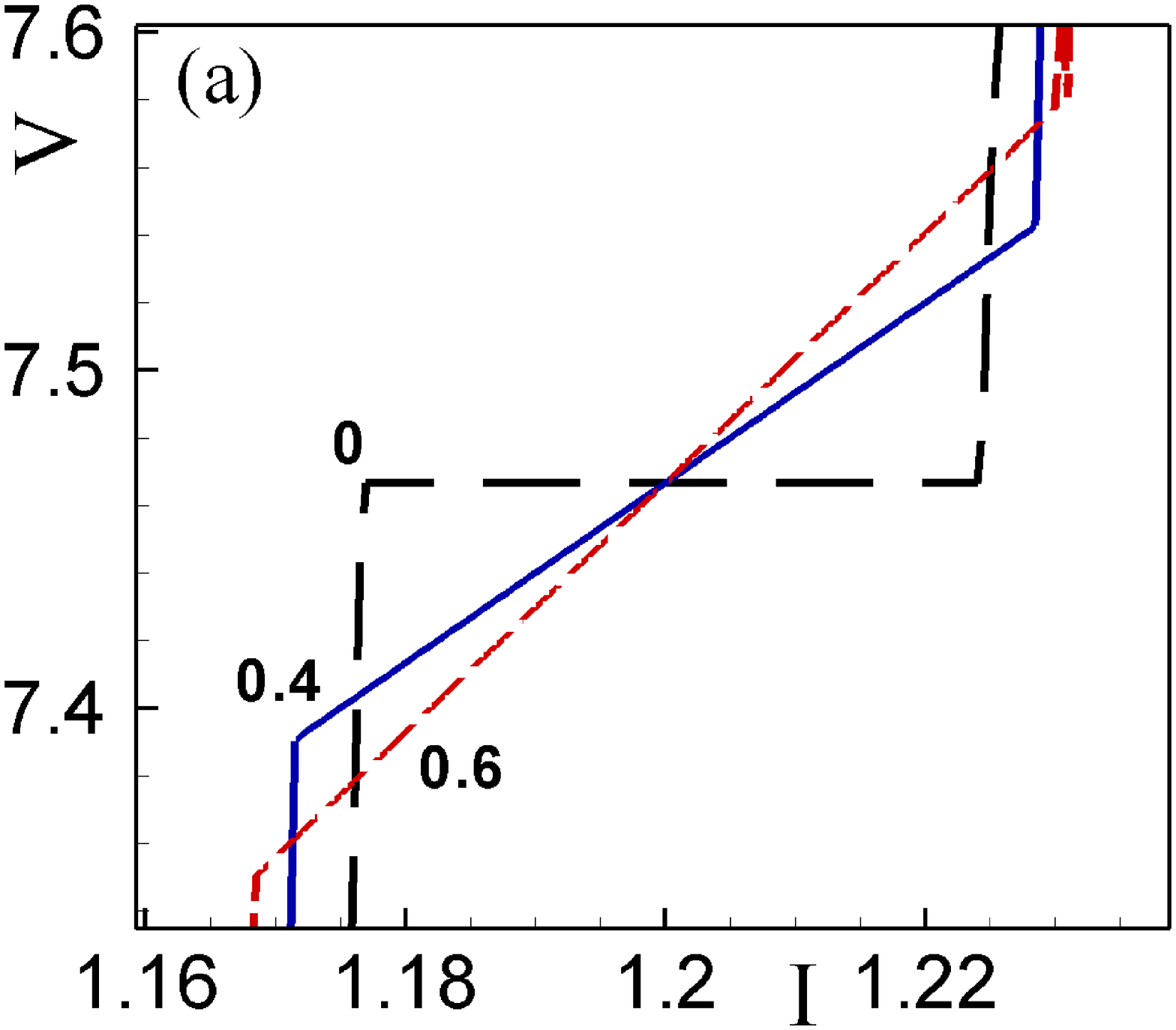} \includegraphics[ width=.4\linewidth, angle =-90]{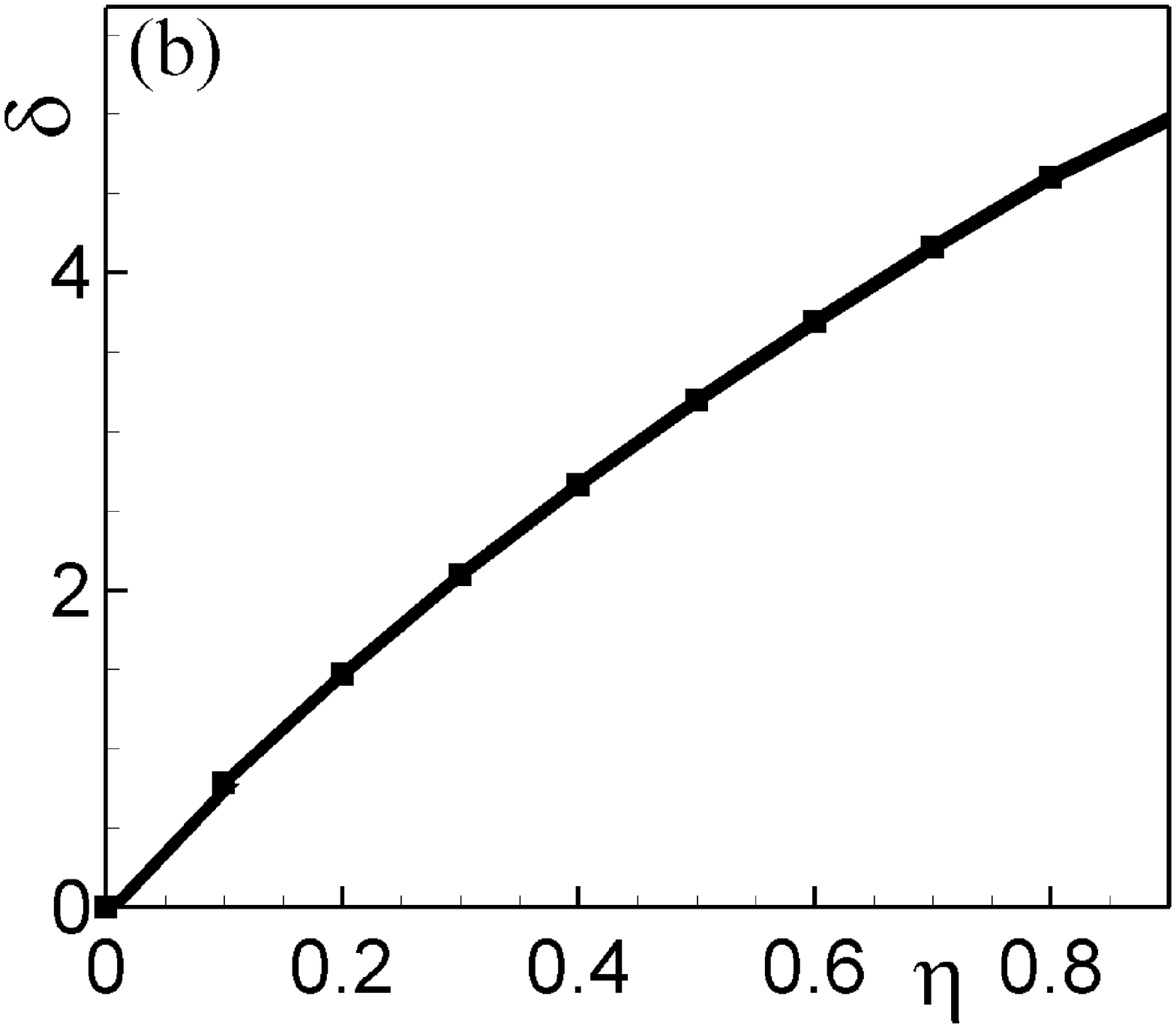}.
	\caption{ (Color online) (a) The enlarged part of the IV-characteristics in the SS region at different values of the nonequilibrium parameter: $\eta=0$ -- thick dashed line, $\eta=0.4$ -- solid line, $\eta=0.6$ -- thin dashed line; (b) The SS slope value for different values of $\eta$. Solid line represents results of fitting.}
	\label{fig3}
\end{figure}
Fig. \ref{fig3}(a) shows that the slope of the Shapiro step $\delta= \Delta V/\Delta I$ increases with $\eta$, while it is absent in the case of $\eta=0$. The slope as a function of $\eta$ shown in Fig.\ref{fig3}(b) demonstrates a monotonic dependence. Fitting of the simulated data (solid line) gives $\delta= 8.18664 \eta - 1.93047 \eta^{2} $. Using the fact that the Debay  screening length  in high temperature superconductors like BSCCO is comparable to the thickness of the S-layer ($r_{D} \approx d_{s}$) we find $\eta=\frac{4\pi r_D\tau_{qp}}{R}$. It brings us to the conclusion that it is possible to estimate the relaxation time for the quasiparticles based on the fitting results.

\begin{figure}[h!]
	\centering	
	\includegraphics[ width=0.4\linewidth, angle =-90]{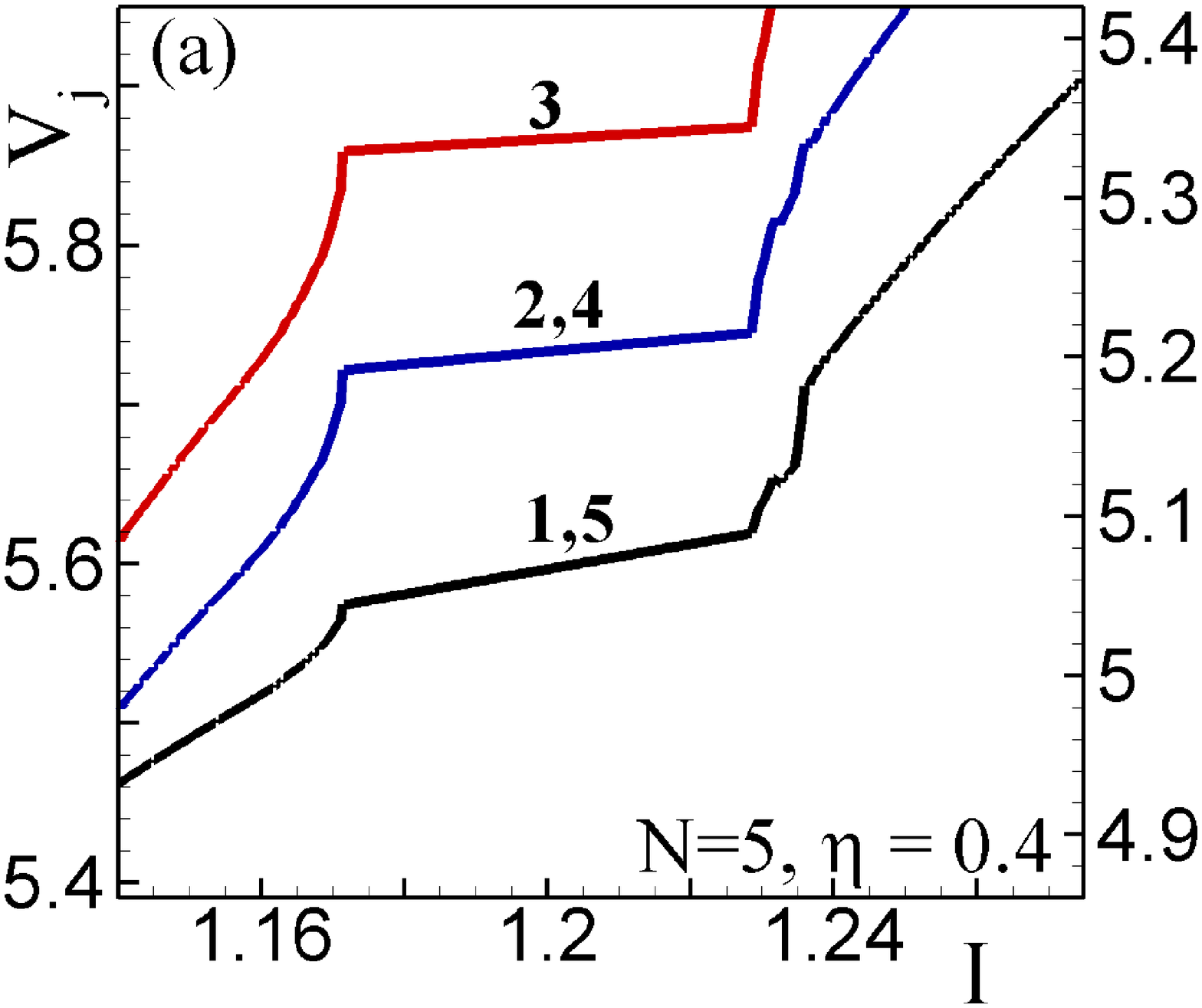}.\includegraphics[ width=.4\linewidth, angle =-90]{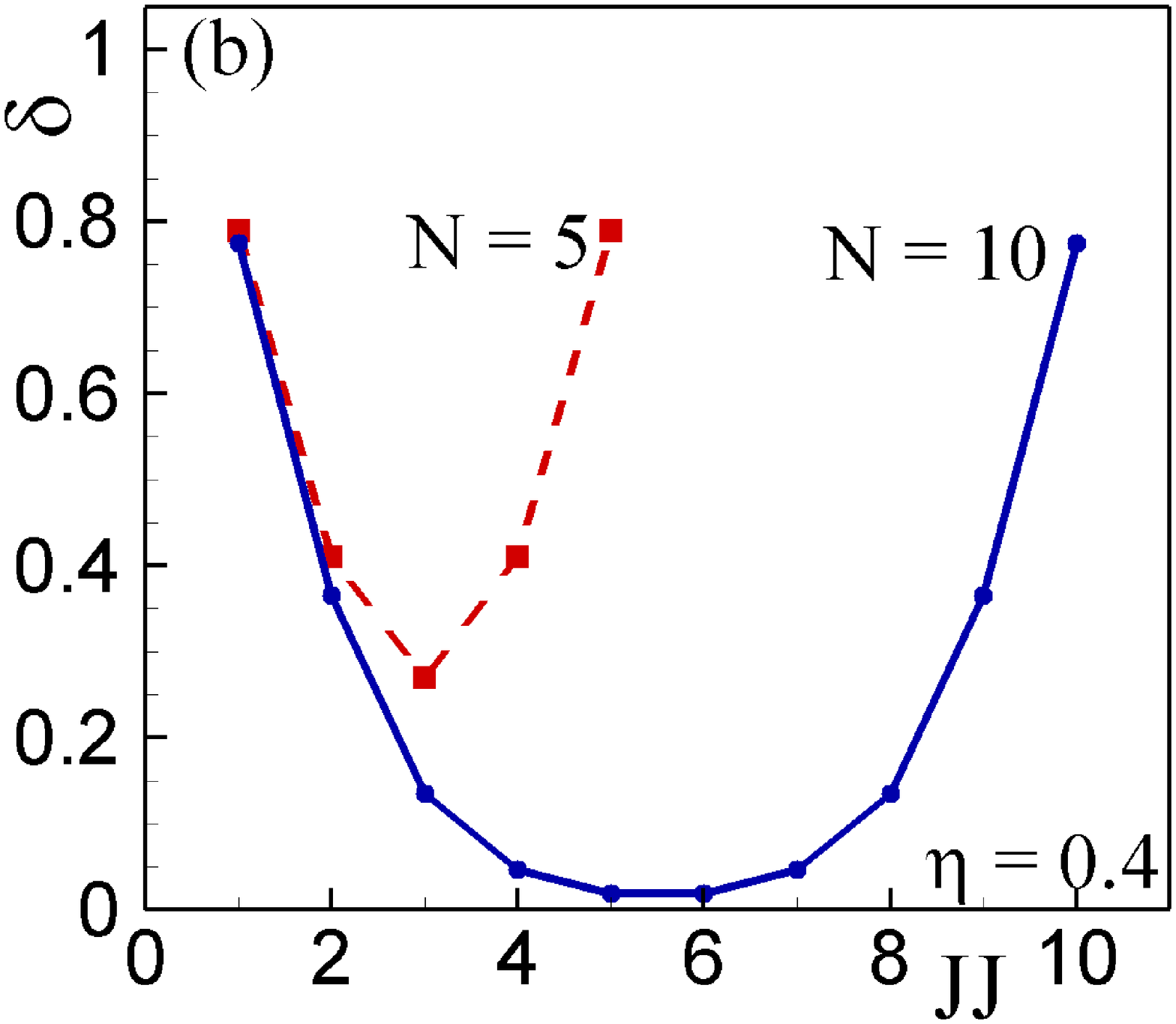}
	\caption { (color online) (a) An enlarged part of the IV-characteristics with the SS for each JJs. Numbers indicate a position of the junction in the stack; (b) Distribution of the SS slope in the stacks with 5 and 10 JJs.}
	\label{fig4}
\end{figure}

There is another interesting feature of the Shapiro step at the nonequilibrium conditions found in the stack of coupled JJs: \emph{the SS slope in the IV-characteristics of each JJ of the stack  can have a different value}. The enlarged parts of the IV-characteristics with the SS for all JJs in the stack  for the case $N=5$ and $\eta=0.6$  are shown in Fig.\ref{fig4}(a).
Additionally to the corresponding shift due to the nonperiodic BCs,  we see here that  the SS of JJ in the middle of the stack has a minimal slope. This feature is related to the fact that the charge imbalance potential at the boundary of the stack has the highest value. The distribution of the SS slope along the stacks with 5 and 10 JJs  is demonstrated  in Fig.\ref{fig4}(b). It shows that with an increase in the number of junctions in the stack, the effect of the charge imbalance on the middle junctions is getting weaker. Particularly, the slope of the second and third JJs in the stack with ten junctions is smaller than in case of $N=5$. Based on this result, it is natural to suppose that in the stack with a large number of JJs the middle ones have practically no slope.

Actually, a finite slope of the Shapiro steps in the IV-characteristics of intrinsic JJs is manifested in some experimental results. Particularly, in Ref.\cite{bae08}  the authors explained it as a manifestation of the phase-diffusion effect. According to our presented results, the slope of the SS might be related to the charge imbalance effect. We note also that the width of the SS increases with the nonequilibrium parameter, as we can see also in the upper inset to Fig.\ref{fig:2}.

Let us now discuss a possibility of the experimental testing of the predicted charge imbalance manifestations in the intrinsic Josephson junctions. To estimate the corresponding values of the parameters, we present the existing experimental data \cite{Ryndyk,yurgens2000intrinsic,Krasnov,Tamasaku} in the Table \ref{tab:table1}.
\begin{table}[h!]
	\caption{\label{tab:table1} Parameter values for the intrinsic Josephson junctions in high temperature superconductors}
	\begin{ruledtabular}
		\begin{tabular}{lcr}
		Parameter & Value & for estimations\\
		\hline
		N(0), $states/eV cm^{3}$ & $10^{22}$ & $ 10^{22}$ \\
		$d_{s}, $ $\AA{}$ & $ 3 \sim 5 $ &  4 \\
		$\Delta(T)$, $meV $& from $30$ to 0 at T=$T_{c}$ & 20 \\
		$I_{c}$, A/cm$^{2}$ & 10$^{2}$ $\sim$ $10 ^{5}$ & 10$^{4} $ \\
		$\tau_{qp}$, ps & $1$ $\sim$  $1000$ at T=4.2 K & 300  \\
		[.1ex]
		\end{tabular}
	\end{ruledtabular}
\end{table}
Particularly, we can estimate the tunneling frequency $\nu$ for the quasiparticles\cite{Ryndyk} based on the formula $\nu = \frac{I_{c}(0)}{2 \pi e \Delta N(0) d_{s}}$, which gives $\nu \approx $ 1.24 x 10 $^{9}$ s$^{-1}$. It allows one to find the nonequilibrium parameter  $\eta =\nu \tau_{qp}=0.375$. In principal, the nonequilibrium parameter can be larger at the corresponding choice of the parameter values presented in the table. For the normalized relaxation time we have obtained $\zeta= \omega_{p} \tau_{qp}\sim 0.3$  at $\omega_{p} \sim 1 GHz$. So, we expect at the estimated values of the relaxation time and nonequilibrium parameter that the experimental IV-characteristics would clearly demonstrate a slope and the shift of the Shapiro step.

The effect of the charge imbalance on the Shapiro step in the first branch of the IV-characteristic is experimentally studied in Ref.\cite{rother03}. The observed shift of the Shapiro step from the canonical value is explained by the charge imbalance in the superconducting layer close to the normal electrode. The corresponding JJ was in the resistive state. Taking into account the contact voltage between the normal electrode and the first S-layer\cite{Ryndy-kkell,rother03}, the authors have determined the new position $V^{*}_{ss}$ of the SS by
\begin{equation}
V^{*}_{ss} = \frac{\hbar\omega}{2 e} -\delta V
\end{equation}
where $\delta V=J \tau_{qp}/(2 e^2 N(0))$.  The theory of the stationary charge imbalance effect ($dv/dt=0$) was used for the explanation of the experimental results when the charge imbalance potential on the S layer was determined by

\begin{equation}
\Psi_{n} =\frac{\tau_{qp}}{2 e^{2} N(0)} \left( J^{qp}_{l-1}-J^{qp}_{l}\right)
\label{eq:stationary}
\end{equation}

We note that based on the value of the SS shift, we can determine the relaxation time for the quasiparticles  as
\begin{equation}
\tau_{qp} =\delta V \frac{A 2 e^{2}N(0)}{I}
\end{equation}
where A is the area of the mesa and $I$ is the biased current.

In general, the intrinsic JJs are in the nonstationary state at any value of the bais current\cite{Dmitry}, and the effect of the charge imbalance on the SS in this case is not investigated yet. The dynamics of the quasiparticle potential is now determined by kinetic equations (7-9) instead of equation (\ref{eq:stationary}). As we have demonstrated, the nonstationary charge imbalance leads to a slope in the SS. The slope and the width of the SS depend on the value of the nonequilibrium parameter. Probably, the slope of SSs is manifested in the experimental results of Ref.\cite{bae08}, but the authors explained it as a result of phase diffusion. Also, we can see a small slanting in the results of Ref.\onlinecite{rother03} (see Fig.3  there). The answer to the question how strong the nonequilibrium effects  are in the system can be obtained by measurements of the Shapiro steps slope.

We note the importance of the role of boundary and proximity effects in intrinsic Josephson junctions causing the nonperiodic boundary conditions which are not investigated carefully yet. As we demonstrated in the present paper, the nonperiodic boundary conditions can be a reason for the SS shift in the experimental IV-characteristics.

	As summary, we have investigated the effect of the charge imbalance on the Shapiro step in the outermost branch at the nonequilibrium conditions. Two important features for the Shapiro step are predicted. First, the Shapiro step demonstrates a shift of its position from the canonical value $N \omega$, where $N$ is the number of junctions in the stack and $\omega$ is the frequency of the external radiation. The value of this shift depends on the boundary conditions and coupling between Josephson junctions. Due to the coupling, the effect of the boundary conditions is extended to the neighboring junctions. Second, the Shapiro step demonstrates a finite slope in the IV-characteristics of a stack of coupled junctions. The value of the slope depends on the value of the nonequilibrium parameter. The origin of the slope is related to the charge imbalance in the superconducting layers because it is absent in the resistively and capacitively shunted junction models.

The authors thank M. Kupriyanov, V. Krasnov, V. Ryazanov, and I. Rahmonov for fruitful discussions. This work is supported by RFBR grant 15--51--61011 and JINR - EGYPT collaboration in 2015. The authors thank T. Hussein, Kh. Hegab and D. Kamanin for support of this work.

\end{document}